\documentclass[useAMS,usenatbib]{mn2e}
\usepackage{psfrag,graphicx}
\usepackage{epstopdf}
\usepackage{txfonts}
\newcommand{\atlas}{ATLAS$^{\mathrm{3D}}$}
\newcommand{\hi}{{\sc H\,i}}
\title[Star formation in the outer regions of NGC~4203]{Star formation in the outer regions of the early-type galaxy NGC~4203}
\author[M.K. Y{\i}ld{\i}z et al.]{Mustafa K. Y{\i}ld{\i}z$^{1,2,3}$\thanks{E-mail: mkyildiz@astro.rug.nl}, Paolo Serra$^{3}$, Tom A. Oosterloo$^{1,4}$,Reynier F. Peletier$^{1}$, \and Raffaella Morganti$^{1,4}$, Pierre-Alain Duc$^{5}$, Jean-Charles Cuillandre$^{5,6}$, Emin Karabal$^{5,7}$\\\\
$^{1}$Kapteyn Astronomical Institute, University of Groningen, P. O. Box 800, 9700 AV Groningen Netherlands\\
$^{2}$Astronomy and Space Sciences Department, Science Faculty, Erciyes University, Kayseri, Turkey\\
$^{3}$CSIRO Astronomy and Space Science, Australia Telescope National Facility, PO Box 76, Epping, NWS 1710, Australia\\
$^{4}$Netherlands Institute for Radio Astronomy (ASTRON), Postbus 2, 7990 AA Dwingeloo, The Netherlands\\
$^{5}$Laboratoire AIM Paris-Saclay, CEA/Irfu/SAp – CNRS – Universit´e Paris Diderot, 91191 Gif-sur-Yvette Cedex, France\\
$^{6}$Observatoire de Paris, PSL Research University, France\\
$^{7}$European Southern Observatory, Karl-Schwarzschild-Str. 2, 85748 Garching, Germany}

\begin{document}

\date{Submitted 0 August 1234; Accepted 0 October 1234}
\maketitle

\label{firstpage}
\begin{abstract}
{NGC 4203 is a nearby early-type galaxy surrounded by a very large, low-column-density \hi \ disc. In this paper we study the star formation efficiency in the gas disc of NGC~4203 by using the UV, deep optical imaging and infrared data. We confirm that the \hi \ disc consists of two distinct components: an inner star forming ring with radius from $\sim$1 to $\sim$3 R$_{eff}$, and an outer disc. The outer \hi \ disc is 9 times more massive than the inner \hi \ ring. At the location of the inner \hi \ ring we detect spiral-like structure both in the deep $g'-r'$ image and in the 8~$\mu$m \textit{Spitzer}-IRAC image, extending in radius up to $\sim$~3 R$_{eff}$. These two gas components have a different star formation efficiency likely due to the different metallicity and dust content. The inner component has a star formation efficiency very similar to the inner regions of late-type galaxies. Although the outer component has a very low star formation efficiency, it is similar to that of the outer regions of spiral galaxies and dwarfs. We suggest that these differences can be explained with different gas origins for the two components such as stellar mass loss for the inner \hi \ ring and accretion from the inter galactic medium (IGM) for the outer \hi \ disc. The low level star formation efficiency in the outer \hi \ disc is not enough to change the morphology of NGC~4203, making the depletion time of the \hi \ gas much too long.}

\end{abstract}
\begin{keywords}
galaxies: elliptical and lenticular -- galaxies: evolution -- galaxies:individual(NGC~4203) -- galaxies:ISM -- galaxies:structure
\end{keywords}

\section{Introduction}
\label{sec:introduction}

Although we have long known that early-type galaxies (ETGs) can harbour a stellar disc, their actual distribution of bulge-to-disc ratios B/D has been debated for many decades \citep[e.g.,][]{1951ApJ...113..413S, 1970ApJ...160..831S, 1976ApJ...206..883V}. Recently, a number of studies based on broad-band imaging and integral-field spectroscopy at optical wavelengths have concluded that most ETGs host a stellar disc, and that ETGs as a family cover a similar range of B/D ratio as spirals \citep{2011MNRAS.413..813C, 2011MNRAS.418.1452L, 2012ApJS..198....2K, 2013MNRAS.432.1768K,2014MNRAS.444.3340W}. Discs are detected also with a number of additional techniques such as narrow-band imaging of emission lines (e.g.,  H$\beta$), imaging of dust absorption, or the observation of the kinematics of planetary nebulae and globular clusters \citep{2003Sci...301.1696R,2006MNRAS.366.1151S,2013MNRAS.428..389P}. Taken together, these results suggest that galaxy evolution is dominated by processes which allow the growth and survival of stellar discs in galaxies of both early- and late type.

Discs have also been found in the cold gas phase of ETGs by means of both neutral hydrogen (\hi) and CO observations \citep[e.g.,][]{1991A&A...243...71V, 2002AJ....124..788Y, 2006MNRAS.371..157M, 2007A&A...465..787O, 2010MNRAS.409..500O, 2011MNRAS.410.1197C}. More recently, \atlas \ observations of \hi \ and CO have shown that $\sim$50 percent of all ETGs contain some cold gas \citep[hereafter S12]{2011MNRAS.414..940Y, 2012MNRAS.422.1835S}. Molecular gas is typically found in small gas discs in the central regions and is linked to small amounts of star formation \citep{2013MNRAS.432.1796A,2013MNRAS.429..534D}. In contrast, in $\sim$20 percent of ETGs outside clusters \hi \ has been found distributed in low-column density discs or rings with typical sizes of many tens of kpc, much larger than the stellar body (S12). If the column density of the gas is high enough, these \hi \ discs are potential fuel for star formation (SF). Contrary to the general idea that galaxies evolve from late- to early type, these \hi \ reservoirs could generate the growth of stellar discs and therefore trigger the transition of the host from an early- to a late-type morphology \citep{2009MNRAS.400.1225C, 2012ApJ...761...23F, 2012ApJ...755..105S}. 
\begin{figure*}
\includegraphics{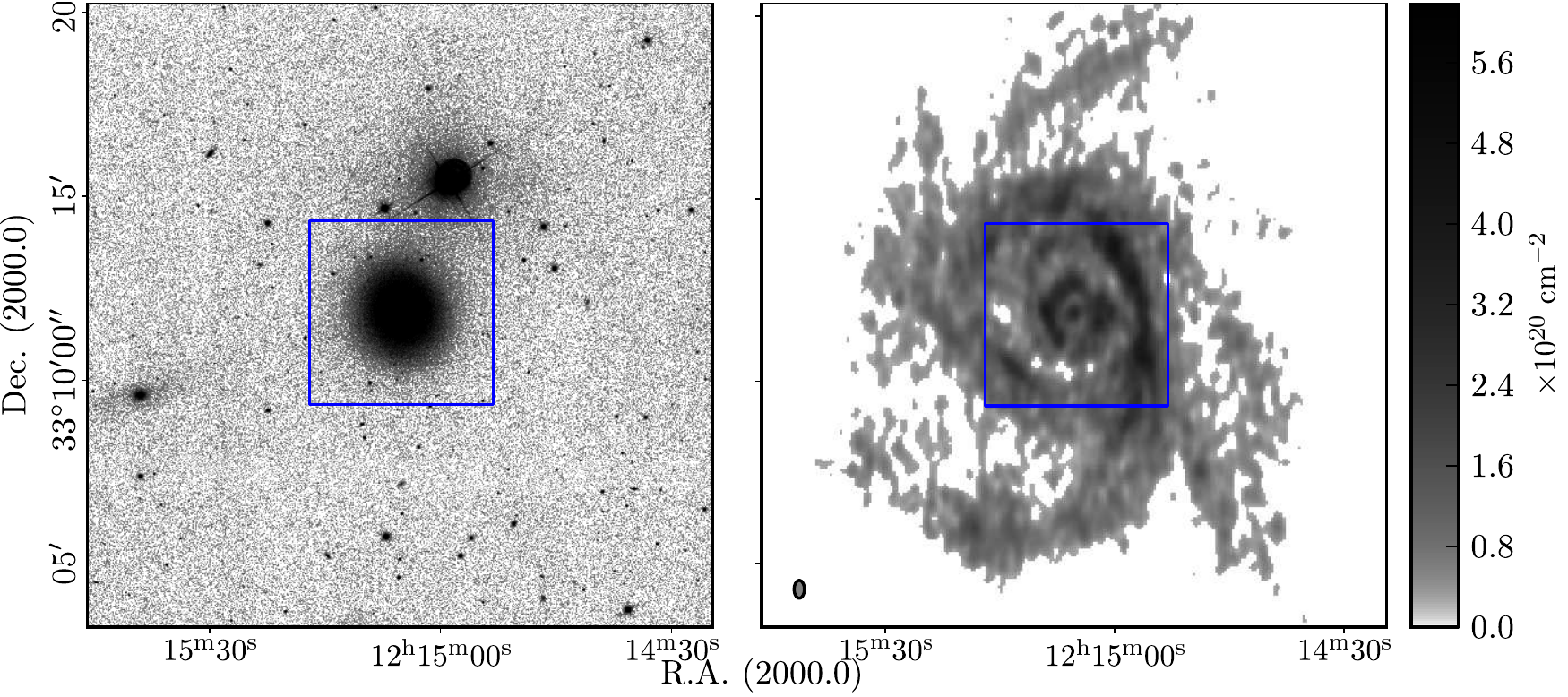}
\caption{SDSS g-band(left) and new deep WSRT \hi \ (right: see Sec. \ref{sec:wsrt}) images of NGC~4203. The size of the images is $\sim74\times74$kpc$^2 \ (17'\times17')$. The colourbar refers to the \hi \ image. The \hi \ beam is shown in the bottom-left corner of the \hi \ image. The blue boxes represent the size of the images shown in Fig. \ref{fig:images2}.}
\label{fig:images}
\end{figure*}

In this context, many recent studies have focused on the presence of SF in the outer discs of  ETGs \citep{2010ApJ...714L.290S, 2011ApJ...733...74L, 2012ApJ...745...34M, 2012ApJ...755..105S}. Such SF is typically traced by UV light emitted by massive stars of age up to several hundreds Myr \citep{2007ApJS..173..357K, 2011Ap&SS.335...51B, 2012ApJ...755..105S}. \citet{2012ApJ...745...34M} show that $\sim$40 percent of 38 red sequence and blue cloud ETGs with stellar mass below $\sim 4\times10^{10}$ M$_{\odot}$ have extended UV discs, which is evidence for disc growth in these objects. 

In this paper, we aim to bring these two lines of investigation (\hi\ and UV) together and study SF in the large \hi \ disc of the early-type galaxy NGC~4203. This is a nearby, nearly face-on lenticular. Figure \ref{fig:images} shows an SDSS $g$-band image of the galaxy (left panel) and Table \ref{table:parameters} provides some basic information. NGC~4203 has been observed at many wavelengths. Optical studies show that NGC~4203 has a relatively massive stellar disc \citep[$B/D=0.4$-0.5 --][]{1979ApJ...234..435B,2013MNRAS.432.1768K}. Early \hi \ observations revealed a low-column-density \hi \ disc with a complex morphology \citep[][]{1981ApJ...250..517B,1988A&A...191..201V}. More recent observations show that the \hi \ disc extends to very large radius and is characterised by spiral-like \hi \ arms (S12). This gas rotates regularly around the galaxy and is kinematically aligned with the stellar disc \citep[][]{2014MNRAS.444.3388S}.
\begin{table}
\begin{center}
\caption{General properties of NGC~4203}
\begin{tabular}{lcccc}
\hline
\hline
Quantity	& Value	& Unit	& Reference	& \\
\hline
Hubble Type	& SAB0			& --			& (1)	& \\
R.A. (J2000)	& $12^{h}15^{m}05.06^{s}$			& HMS	& (2)	& \\
Dec. (J2000)	& $+33^{d}11^{m}50.382^{s}$			& DMS	& (2)	& \\
Helio. Rad. Vel.	&1086$\pm$4	& km/s		& (1)	& \\
Distance		& 14.7		& Mpc		& (3)	& \\
M$_{K}$			& -23.4	& mag		& (3)	& \\
A$_{B}$			& 0.05	& mag		& (3)	& \\
Effective Radius	& 29.5	& arcsec	& (3)	& \\
log$(M/L)_{star}$	& 0.5		& M$_{\odot}/$L$_{\odot r}$	& (4)	& \\
log$M_{star}$	& 10.6	& M$_\odot$	& (4)	& \\
Galactic $E(B-V)$	& 0.012		& mag		& (5)\\
\hline
\end{tabular}
\label{table:parameters}
\end{center}
\hspace{0cm} {\footnotesize
References:(1) \citet{1991rc3..book.....D}; (2) \citet{2005ApJ...627..674A} (3) \citet{2011MNRAS.413..813C}; (4)\citet{2013MNRAS.432.1862C}; (5) \citet{1998ApJ...500..525S}};.
\end{table}

Despite the large \hi \ reservoir ($\sim10^{9}\ \mathrm{M}_{\odot}$), the stellar population is uniformly old ($\sim10\ \mathrm{Gyr}$) inside 1 $R_{\mathrm{eff}}$ \citep{2015MNRAS.448.3484M} suggesting that the \hi \ is not fueling a high level of SF in this region. Yet, there are signs of possible low-level SF within the stellar body. For instance, NGC~4203 contains $2.5\times10^{7}\ \mathrm{M}_{\odot}\ \mathrm{of \ H}_{2}$ in the centre  \citep{2003ApJ...584..260W,2011MNRAS.414..940Y}. Hubble Space Telescope observations show nuclear dust structures in the very centre of the galaxy \citep[300 pc radius;][]{2003ApJS..146..299E}. The dust is visible also on a large scale in a 8 $\mu$m image, where the flux, mostly from polycyclic aromatic hydrocarbons (PAHs) appears distributed in a spiral-like pattern (\citealt{2004ApJS..154..229P}; we will make use of this image as well as of a 24 $\mu$m image in support of our analysis of SF in NGC~4203).

Here we use new, deep \hi \ observations and UV imaging from the \textit{Galaxy Evolution Explorer} (GALEX) to study the relation between \hi \ and star formation as well as the impact of this large gas reservoir on NGC~4203. In Section 2 we describe the data (\hi, FUV, infrared as well as ancillary optical imaging) and the methodology used for our work. In Section 3 we discuss global properties of the \hi, FUV, infrared and optical images, present the derivation of star formation rate (SFR) from the data and compare the SF efficieny in the \hi \ disc with that of late-type and dwarf galaxies. In Section 4 we discuss the origin of the \hi \ gas and its impact on the host galaxy. In Section 5 we give a summary of the work.

\section{Data and analysis}
\label{sec:data_analysis}
\subsection{WSRT data reduction}
\label{sec:wsrt}
We select NGC~4203 from the sample of ETGs with extended \hi \ discs presented by S12. We observe this galaxy for 9$\times$12 hours using the Westerbork Synthesis Radio Telescope (WSRT). Compared to the original 12-h observation by S12, this longer integration allows us to image the \hi \ down to a similar column density but at a significantly higher angular resolution. This is desirable to compare the \hi \ image to the UV image used to study SF (see below).

We reduce separately the 9 \hi \ datasets (12-h integration each) following standard techniques with the MIRIAD package \citep{1995ASPC...77..433S}. Having calibrated and subtracted the continuum from the datasets, we combine them in the UV plane and image with robust=0.4 weighting, obtaining an \hi \ cube with an angular resolution of 29.3$\times$16.7 arcsec$^{2}$ (PA=0.5 deg). The velocity resolution is 16 km s$^{-1}$ after Hanning smoothing. The noise level of the data cube is 0.12 mJy beam$^{-1}$.

We create an \hi \ image by summing flux at all velocities using only pixels included in a mask. In order to include diffuse, faint emission in the mask, we select pixels above a certain threshold in: (i) the original data cube; (ii) four data cubes smoothed spatially with a 2D Gaussian of FWHM 30, 45, 60 and 90 arcsec, respectively; (iii) three data cubes smoothed in velocity with a Hanning filter of width 32, 64 and 112 kms$^{-1}$; (iv) six data cubes obtained smoothing in velocity the 30- and 60 arcsec resolution cubes as in point (iii). We create the mask using a 3$\sigma$ threshold except for the 90-arcsec resolution cube, for which we adopt a 4$\sigma$ threshold. 

\begin{figure*}
\includegraphics{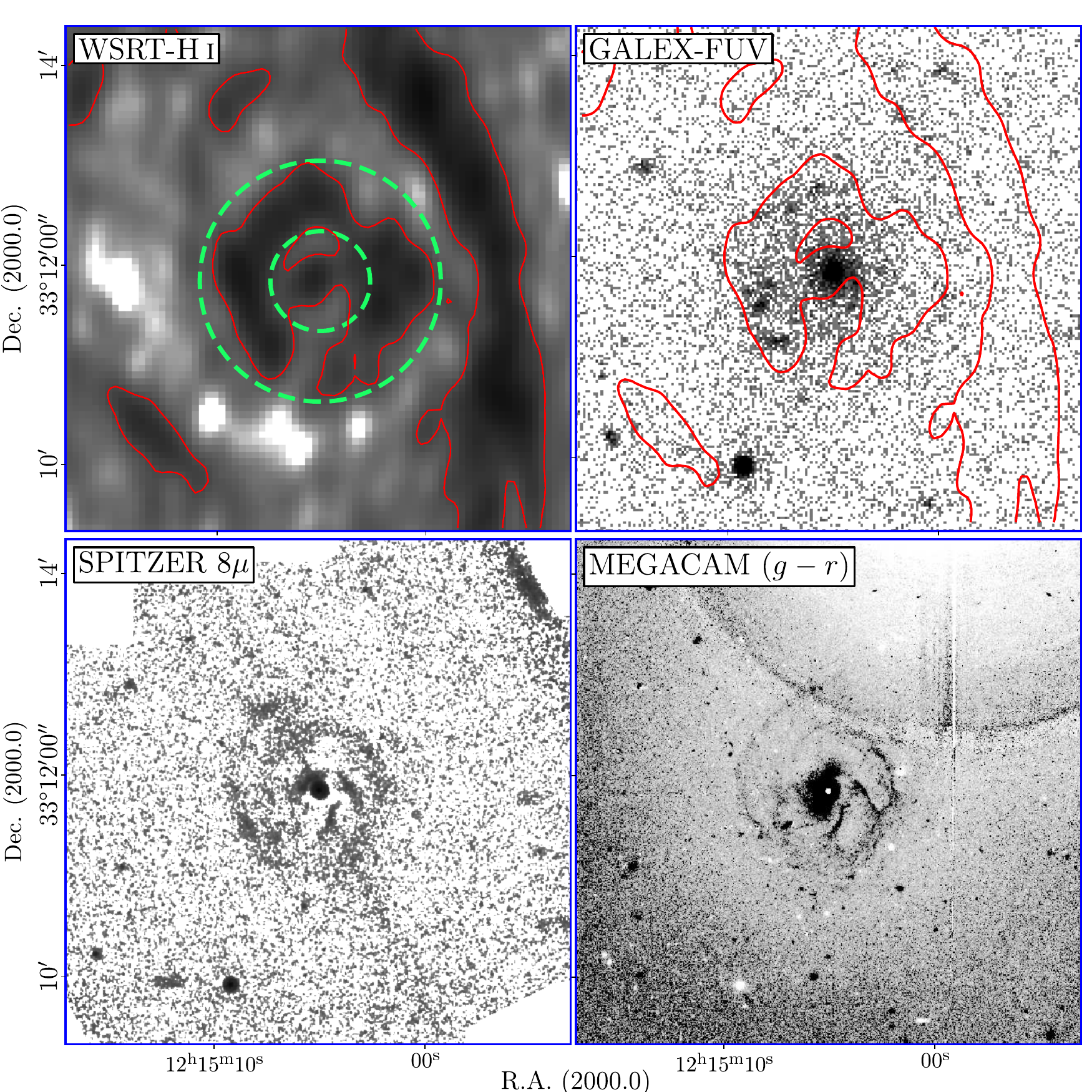}
\caption{Zoomed-in images of the deep \hi \ (top-left), GALEX-FUV (top-right), Spitzer $8 \mu$m infrared (bottom-left) and deep MegaCam $g'-r'$ colour (bottom-right) images. All images show a field of size $300''\times300''$ ($\sim$ 22kpc$\times$22kpc). Dark areas correspond to redder regions in the deep MegaCam colour image. The red contours on the \hi \ and FUV images represent an \hi \ column density of $1.8\times 10^{20}$ cm$^{-2}$. The location of the inner \hi \ ring is shown by the green-dashed lines.}
\label{fig:images2}
\end{figure*}

We determine a detection limit from the original data cube as a 5$\sigma$ signal in a single velocity resolution element, obtaining $2.3\times10^{19}$ cm$^{-2}$. We obtain the final \hi \ image by applying the WSRT primary beam correction to all pixels above this detection limit in the moment zero image derive from the masked \hi \ cube. We show the \hi \ image in Fig. \ref{fig:images} (right panel) and a zoom in on the central 300 arcsec in Fig. \ref{fig:images2} (top-left panel).
\subsection{GALEX data reduction}
\label{sec:galex}

We obtain GALEX UV images of NGC~4203 from the Mikulski Archive for Space Telescopes. We use data from the Calibration Imaging, whose goals and specifications are described in \citet{2005IAUS..216..221M}. The GALEX field of view (FOV) is $\sim$ 1.2 degrees in diameter and the instrument has two main bands: the far-UV (FUV) and the near-UV (NUV). The effective wavelengths for the FUV and NUV channels are 1516\AA \ and 2373\AA, with 4.5 arcsec and 6 arcsec resolution (FWHM), respectively. Details can be found in \citet{2005ApJ...619L...7M, 2007ApJS..173..682M}. Our study is mostly based on the FUV data because the FUV emission is more sensitive to recent star formation than the NUV emission \citep{2011Ap&SS.335...51B}.

Since NGC~4203 has not been observed as a target galaxy, the position of the galaxy is different in all the 10 GALEX images. Therefore, we have selected 5 GALEX images where NGC~4203 is sufficiently far from the edge of the GALEX FOV to grant good image quality and reasonable background uniformity. The total exposure time of these 5 UV images is 1186 sec. For the purpose of comparing the distribution of UV emission and \hi \ gas, we analyse a $1500\times1500$ arcsec$^{2}$ GALEX image centered on NGC~4203. Figure \ref{fig:images2} shows the central 300 arcsec of the FUV image, which is the only region where emission from SF in the \hi \ disc is clearly visible by eye.

We make use of the SDSS photometric catalogue \citep{2012ApJS..203...21A} to mask background and foreground objects based on their type (unresolved or extended), brightness, colour and Petrosian radius. For the unresolved sources we adopt a radius equal to the GALEX point spread function (PSF), otherwise we use the Petrosian radius of the object itself. Following visual inspection of the UV images, we choose to mask the objects with $g'-r'$ $\leq$1 and 1.5, and with $g$ $\leq 19.5$ and 21.1 mag for the unresolved and resolved sources, respectively. We estimate the local background from the masked images by applying a moving mean filter of size $300\times300$ pixel$^{2}$ ($450\times450 \ \mathrm{arcsec}^{2}$). The filtering is performed iteratively by clipping three times at a $4.5\sigma$ level. To avoid over-estimating the background, we mask a large area (300 arcsec diameter) around the centre of the galaxy. We subtract this background from the original, masked image. Additionally, a flat residual background is subtracted from the UV image before proceeding with our analysis (see Sec. \ref{sec:rad_pro}).

Finally, we correct the FUV image for the effect of Galactic extinction. We use an $E$(B-V) = 0.012 based on the dust images by \citet{1998ApJ...500..525S} and assume that $A_{\mathrm{FUV}}$=8.24~$\times~E$(B-V) and $A_{\mathrm{NUV}}$=8.2~$\times~E$(B-V) \citep{2007ApJS..173..293W}. Thus, we use the following equations for the Galactic correction:
\begin{equation}
I_{FUV,corr}= I_{FUV} \times \mathrm{1.1}
\label{eq:eq1}
\end{equation}
\begin{equation}
\label{eq:eq2}
I_{NUV,corr}= I_{NUV} \times \mathrm{1.08}
\end{equation}
\begin{figure*}
\includegraphics{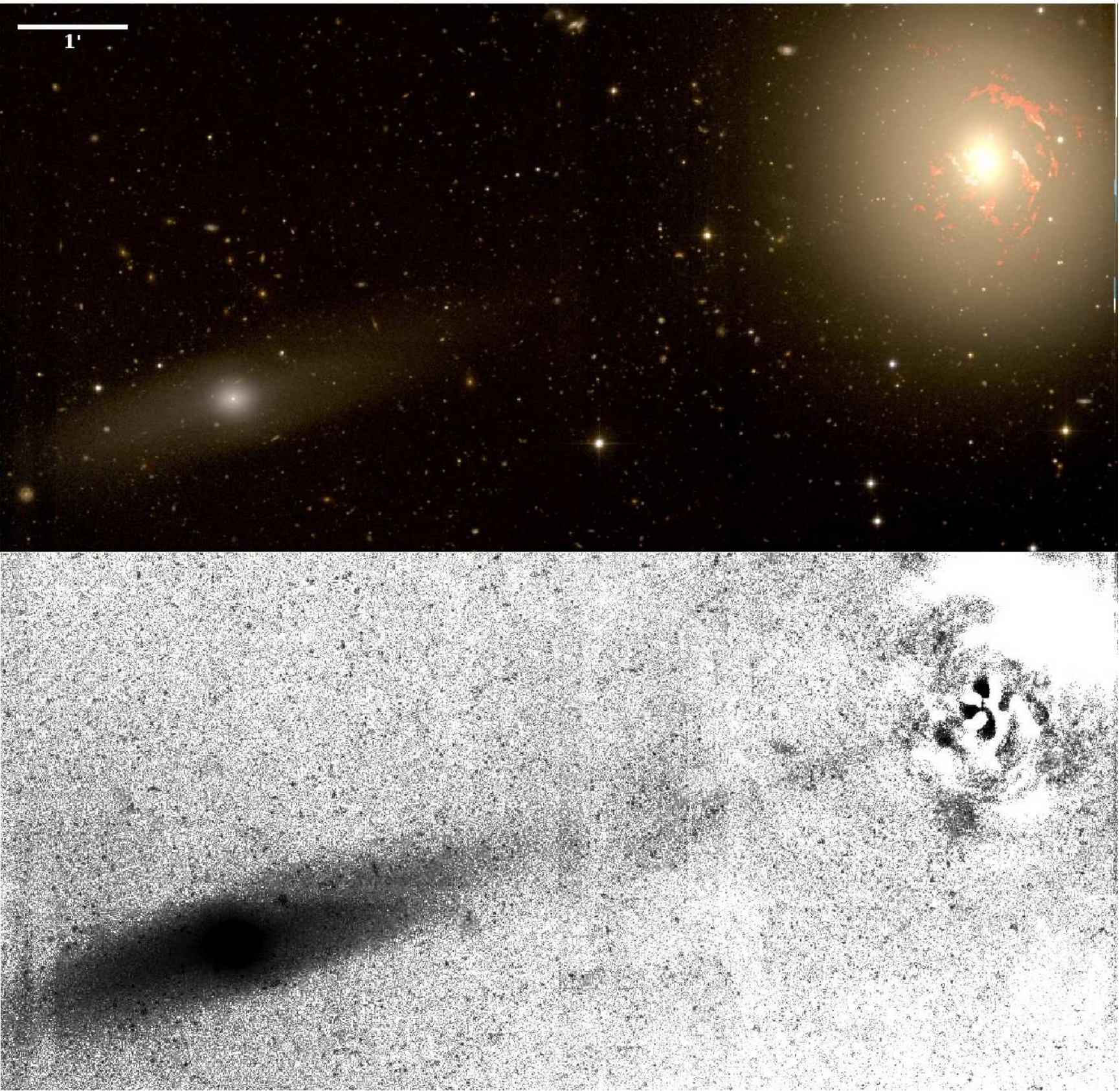}
\caption{\textit{Top}: Composite $g'+r'$ Megacam image of NGC~4203 with the dust lanes superimposed in red. The dust lanes are obtained by summing the residual $g'+r'$ images after a galaxy model subtraction (see Sec. \ref{sec:8mu}). This image also shows the disrupted companion dwarf galaxy at the east of the galaxy. \textit{Bottom}: MegaCam g$'$ band image of the disrupted companion dwarf galaxy and residual map of NGC~4203. It exhibits a possible faint tail East of our galaxy which as the same direction as the two tidal tails of the companion dwarf (see Sec. \ref{sec:propertiesHI}). The white bar shows a lenght of 1 arcmin.}
\label{fig:images3}
\end{figure*}
\subsection{Ancillary optical and infrared images}
\label{sec:8mu}

We support our study of SF, which is primarily based on the UV images described above, with ancillary images at 8 $\mu$m, 24 $\mu$m and optical wavelengths. The 8 $\mu$m mid-infrared emission is generally considered to trace the PAHs heated by UV photons \citep{1984A&A...137L...5L, 1984ApJ...277..623S}. It is thought that 2/3 of the 8 $\mu$m emission from a galaxy originates from heating by stellar populations 100 Myr or younger \citep{2005ApJ...633..871C,2013seg..book..419C}, and that PAH emission is stronger for increasing metallicity and dust content of the interstellar medium \citep[e.g.,][]{2005ApJ...628L..29E, 2010ApJ...715..506M,2013seg..book..419C}. The 24~$\mu$m emission is generally produced by dust following absorption of UV photons emitted by young stars, and traces SF on a time scale of 10 Myr \citep{2005ApJ...633..871C, 2008AJ....136.2782L}.

In this paper we use archival 3.6 $\mu$m, 8 $\mu$m and 24$\mu$m images from \citet{2004ApJS..154...10F} obtained with the Infrared Array Camera and The Multiband Imaging Photometer on the \textit{Spitzer Space Telescope } \citep{2004ApJS..154...25R, 2004ApJS..154....1W}. We remove the background from each image by subtracting the average pixel value calculated in a number of empty regions.

The 8$\mu$m image includes a contribution from stars in the galaxy, which we remove by subtracting the 3.6 $\mu$m image scaled by a factor X$_{3.6}$ as in \citet{2010MNRAS.402.2140S}. These authors find that X$_{3.6}$ has very low scatter about a median value of 0.264 for ETGs, and we adopt this value for NGC~4203. The bottom-left panel of Fig. \ref{fig:images2} shows the resulting 8 $\mu$m-non stellar image. Similarly, the 24~$\mu$m flux is also contaminated by emission from old stellar populations and hot, circumstellar dust \citep[e.g.,][]{2009ApJ...700..161S, 2010ApJ...713L..28K, 2013ApJ...766..121M}. To remove this emission, we apply the same scaling method used for the 8~$\mu$m. We have adopted a scaling factor from \citet{2013ApJ...766..121M} that study the empirical 24~$\mu$m scaling factor for dust-free ETGs.

We also use deep optical images of NGC~4203 that are obtained in the g$'$ and r$'$ bands using the MegaCam camera on the Canada-France-Hawaii Telescope. These observations are obtained as part of the MATLAS / \atlas \ Large Program. The observing strategy and data reduction procedures with the LSB-Elixir pipeline optimized for the detection of extended low surface brightness features are described in  \citet{2015MNRAS.446..120D}. Further processing is carried out to remove the extended artificial halos surrounding two nearby bright stars. 
We determine the stellar PSFs, including their extended wings, with a manual procedure \citep[as explained in ][]{2015MNRAS.446..120D}. On the resulting image (see Fig. \ref{fig:images3}, top), the early type galaxy exhibits a rather regular extended outer stellar halo, and a dwarf galaxy can be seen east of NGC~4203.

We compute a model of the NGC~4203 with the ellipse fitting procedure of IRAF for both the g$'$ and r$'$ band images and then subtract these models from the original images. The resulting residual image together with the g$'$ band image of the dwarf galaxy is shown at the bottom panel of Fig. \ref{fig:images3}, and it is obvious that some extended spiral/ring structures show up in the main body of the lenticular galaxy.

\subsection{SFR calculation}
\label{sec:units}

We estimate the SFR using the various tracers described above (FUV, non-stellar 8~$\mu$m and 24~$\mu$m). In order to convert FUV emission into SFR we use a conversion from \citet{2010AJ....140.1194B},

\begin{equation}
\label{eq:eq3}
\Sigma_{\mathrm{SFR}}[\mathrm{M_{\odot}yr^{-1}kpc^{-2}}] = 0.68\times 10^{-28} \times I\mathrm{_{FUV}[erg\ s^{-1} Hz^{-1} kpc^{-2}]},
\end{equation}

\noindent where I$_{\mathrm{FUV}}$ is the FUV intensity per unit area and the initial mass function (IMF) is assumed to be of a Kroupa type. This estimate could in principle be corrected for internal extinction on the basis of the \hi \ image using the relation $N$(\hi)$/E(B-V)= 5 \times10^{21}$ cm$^{-2}$mag$^{-1}$ presented by \citet{1978ApJ...224..132B}. This relation is derived for the Milky Way and therefore is likely to provide an upper limit on the extinction in metal poor environements. In the case of NGC~4203, the peak \hi \ column density is $\sim 4.5\times10^{20}$~cm$^{-2}$ and, thus, the maximum extinction should be around $\sim$ 25 percent. In the outer regions, where the \hi \ column density is much lower ($\sim 3 \times10^{19}$cm$^{-2}$), the extinction decreases to a negligible $\sim$ 2 percent. In this paper we do not apply this correction but instead trace SF in dusty environments by using infrared imaging.

Although we do not expect a large amount of dust in the outer regions of NGC~4203, Fig. \ref{fig:images2} shows that in the inner regions dust is present and this will affect the calculation of SFR. As mentioned in Section \ref{sec:8mu}, the FUV emission absorbed by dust is re-emitted at infrared wavelengths. We use the conversion from \citet{2005ApJ...632L..79W} to estimate SFR from the 8 $\mu$m non-stellar image:

\begin{equation}
\label{eq:eq4}
\Sigma_{\mathrm{SFR}}[\mathrm{M}_{\odot}\mathrm{yr^{-1}kpc^{-2}}] = 8.43 \times 10^{-3} {I}_{8 \mu m}(\mathrm{non \ stellar)},
\end{equation}
\noindent where I$_{8 \mu m}$ is the non-stellar 8~$\mu$m intensity in MJy~ster$^{-1}$.

We also combine the FUV and non-steallar 24 $\mu$m images to estimate the SFR as in \citet{2008AJ....136.2782L}:

\begin{equation}
\label{eq:eq5}
\Sigma_{\mathrm{SFR}}[\mathrm{M_{\odot}yr^{-1}kpc^{-2}}] = 8.1\times 10^{-2}I_{\mathrm{FUV}} + 3.2\times 10^{-3}I_{24\mu m}(\mathrm{non \ stellar)},
\end{equation}
\noindent where $I_{\mathrm{FUV}}$ and $I_{\mathrm{24}}$ are in MJy~ster$^{-1}$, and similar to Eq. \ref{eq:eq3} a Kroupa type IMF is used.

Below we study the SF at the original resolution of the FUV and infrared images as well as at the resolution of the \hi \ image in order to compare the distribution of neutral gas and SFR. 
\begin{figure*}
\includegraphics{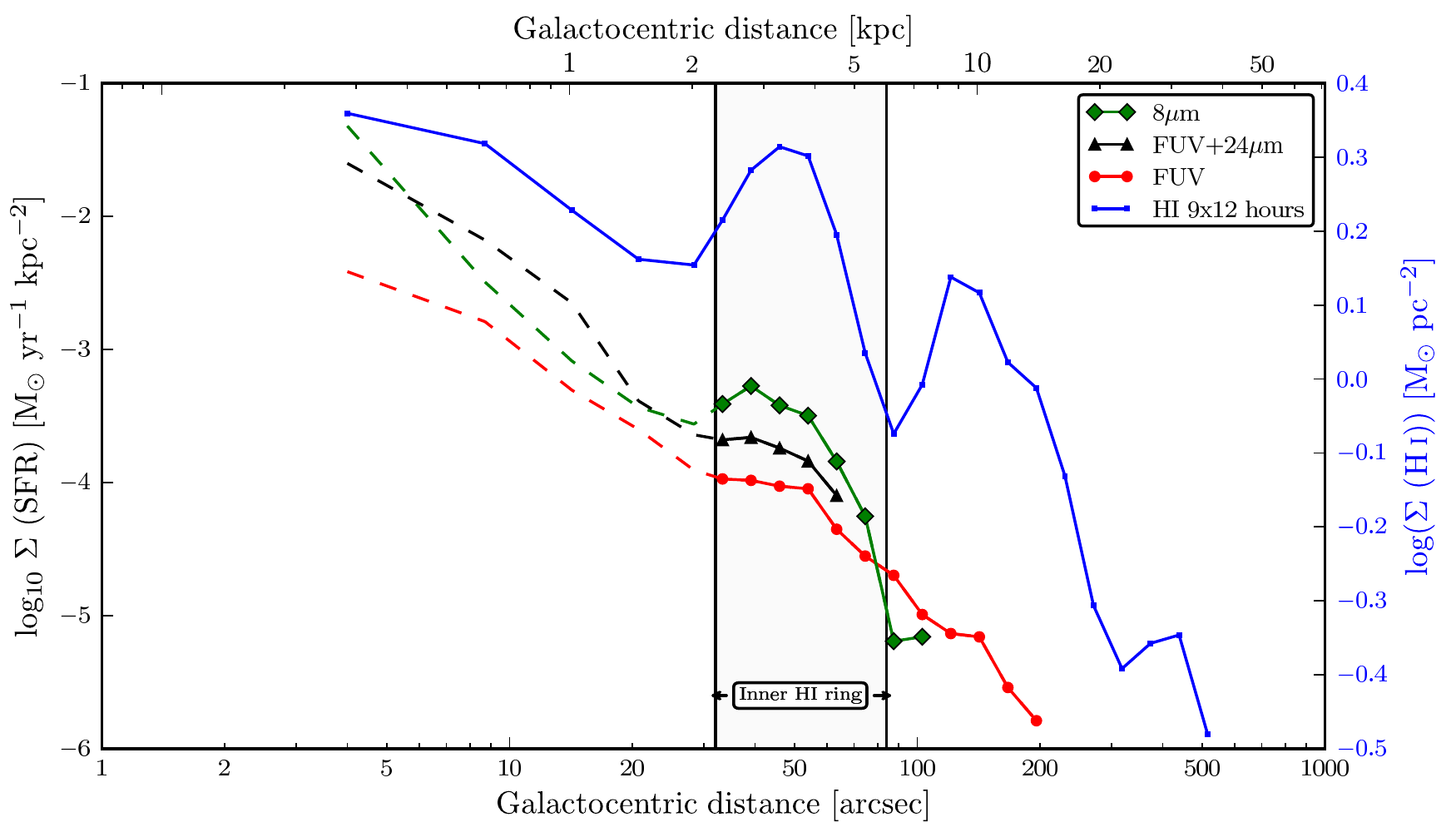}
\caption{Comparison of different SFR radial profiles together with the \hi \ profile. We calculate the SFR with Eqs. \ref{eq:eq3}, \ref{eq:eq4} and \ref{eq:eq5}. Although, we show the central emission from all the bands, we do not include these emissions in our SFR calculation. The dashed part of the profiles traces emission from the bulge and this is not included the calculation of SFR in the rest of paper. The gray shaded area and \hi \ profile are the same as in Fig. \ref{fig:rp_log}.}
\label{fig:spitandfuv}
\end{figure*}
\section{Global Properties of FUV, \hi, and ancillary images}
\label{sec:properties}
\subsection{Flux distribution of the images}
\label{sec:propertiesHI}
Early studies emphasize that the \hi \ properties of NGC~4203 are similar to those of spiral galaxies. For example, \citet{1981ApJ...250..517B} and \citet{1988A&A...191..201V} find a distorted rotation curve from the \hi \ velocity field, possibly suggesting an oval distortion or a kinematical warp \citep{1978PhDT.......195B}. They find that the \hi \ disc is 3 times larger than the optical body and this is confirmed later by S12 using deeper data.

As described in Section \ref{sec:data_analysis}, our new \hi \ image has similar sensitivity as that of S12 but a much better resolution. We detect \hi \ down to a column density of $2.3\times10^{19}$ cm$^{-2}$ at a resolution of 29.3 $\times$ 16.7 arcsec$^{2}$. Using the new \hi \ observations, one can easily see a dense gas ring whose size is similar to that of the stellar body (Fig. \ref{fig:images}). This ring is embedded in a much larger \hi \ disc characterised by long spiral-like arms. The \hi \ disc extends out to $\sim$ 40 kpc from the galaxy centre and contains $\sim$ 9 times more \hi \ gas than the inner ring. Table \ref{table:parameters2} lists some basic galaxy properties derived from the \hi \ image.
\begin{table}
\begin{center}
\caption{Observational results for NGC~4203}
\begin{tabular}{lcccc}
\hline
\hline
Quantity	& Value	& Unit	& $\pm^{\mathrm{b,c,d}}$	& \\
\hline
log$M$(\hi)	& 9.12				& M$_\odot$		& 2 $\%$	& \\
log$M$(\hi)$_{Ring}$	& 8.12			& M$_\odot$		& 5 $\%$	& \\
log$M$(\hi)$_{Disc}$	& 9.07			& M$_\odot$		& 2 $\%$	& \\
log$[M$(\hi)/$M_{stellar}]$	& -1.442			& --	& --	& \\
FUV$^{a}$		& 16.10				& ABmag			& 0.05	& \\
FUV(R$<$100")		& 16.92				& ABmag			& 0.06	& \\
SFR$_{Ring,8\mu m}$		& 8.4$\times$10$^{-3}$		& M$_\odot$ yr$^{-1}$	& 3.0$\times$10$^{-3}$	& \\
SFR$_{Ring,FUV+24\mu m}$	& 4.3$\times$10$^{-3}$		& M$_\odot$ yr$^{-1}$	& 2.0$\times$10$^{-3}$	& \\
SFR$_{Ring,FUV}$			& 3.8$\times$10$^{-3}$		& M$_\odot$ yr$^{-1}$	& 1.0$\times$10$^{-3}$	& \\
SFR$_{Disc}$		& 9.0$\times$10$^{-4}$		& M$_\odot$ yr$^{-1}$	& 4.0$\times$10$^{-4}$	& \\
SFR$_{Ring+Disc,FUV}$	& 4.7$\times$10$^{-3}$		& M$_\odot$ yr$^{-1}$	& 1.0$\times$10$^{-3}$	& \\
$\overline{\Sigma}SFR_{Ring+Disc,FUV}$	& 5.9$\times$10$^{-5}$& M$_\odot$ yr$^{-1}$kpc$^{-2}$	& 1.3$\times$10$^{-5}$	& \\
\hline
\end{tabular} \\
\label{table:parameters2}
\end{center}
\hspace{0cm} {\footnotesize Note: \\
$^{a}$ We take into account the Galactic extinction correction for all the values below. Look at the text for the details.\\
$^{b}$ The errors are in the same unit as the values.\\
$^{c}$ The errors in \hi \ mass values are derived from the noise level (1$\sigma$) of the mask file that is used to create the \hi \ image. We use an annulus that has the same size of the \hi \ beam.\\
$^{d}$ The errors in the SFR and FUV magnitude values are calculated from the uncertainties in the photometric measurements of the FUV, 8~$\mu$m and 24~$\mu$m images \citep{2007ApJ...666..870C,2007ApJS..173..682M,2008AJ....136.2782L}.}
\end{table}
\subsection{Impact of the \hi \ on the host galaxy}
\label{sec:discussion_3}

Figure \ref{fig:images2} shows that NGC~4203 hosts some diffuse FUV emission at the location of the inner \hi \ ring. Additionally, at the same location, we see spiral-like structure in both the 8~$\mu$m and $g'-r'$ images, which is clumpy and almost identical in these images. Moreover, the extended spiral/ring structures in the model subtracted deep optical image (bottom panel of Fig. \ref{fig:images3}) correspond to the dust lanes visible in the MegaCam $g'-r'$ colour image.

In contrast, the bright \hi \ arm $\sim$ 2 arcmin west of the galaxy (visible in the same figure and referred to as western \hi \ arm in the rest of this paper) does not exhibit any clear FUV emission. A similar picture is evident from the 8~$\mu$m-non stellar image and the $g'-r'$ image. While there is clear emission from PAHs and dust absorption in the \hi \ ring, these features are absent from the region occupied by the western \hi \ arm. The question is why these two regions have such different levels of the FUV and 8~$\mu$m emission despite the very similar \hi \ column density. We will come back to this point in Sec \ref{sec:discussion_1}. 

As indicated in Sec. \ref{sec:8mu}, the deep optical images show a dwarf satellite $\sim$ 7.2  arcmin (32~kpc) east of the NGC~4203. This dwarf is red and at a similar redshift with $V_{helio}=1067\pm24$ kms$^{-1}$ \citep{2007ApJS..172..634A}. The stellar mass of the dwarf is roughly $\sim10^{8}$ M$_{\odot}$ from the SDSS-$i$ band photometry.

The bottom panel of Fig. \ref{fig:images3} reveals prominent tidal tails associated to the dwarf and a faint stream, which has the same direction as these tidal tails. From this point, it is clear that the satellite dwarf galaxy is interacting with NGC~4203, for example, these features may be associated with a single on-going minor merger event. We will discuss this point in Sec. \ref{sec:discussion_2}.

\subsection{\hi, FUV and infrared radial profiles}
\label{sec:rad_pro}

We derive the radial profile of \hi, UV and infrared emission along circular annuli whose width increases logarithmically from inside out. The inner region of the radial profiles (1 $R_{\mathrm{eff}}$) is contaminated by old stars in the bulge and, therefore, we do not include it in our SFR calculation.

\begin{figure*}
\includegraphics{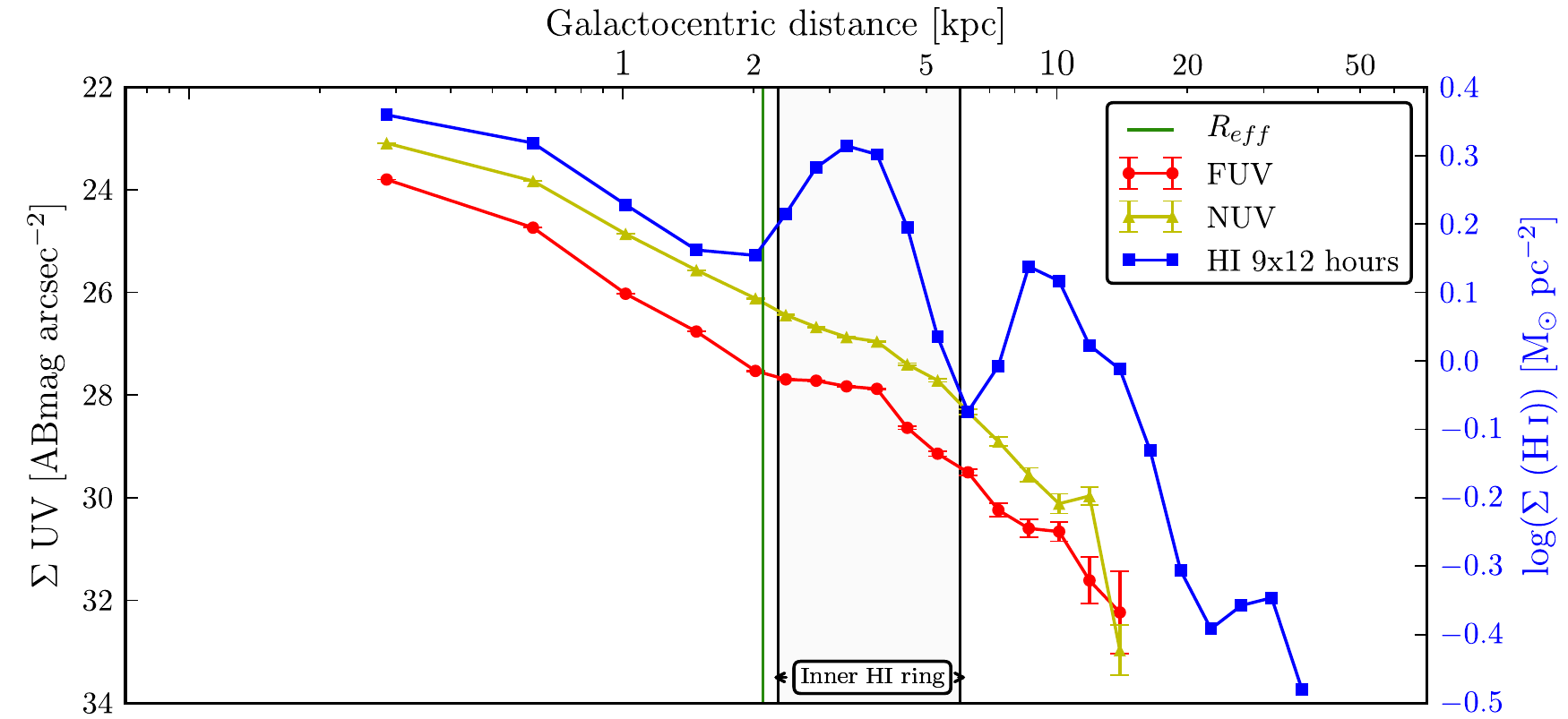}
\includegraphics{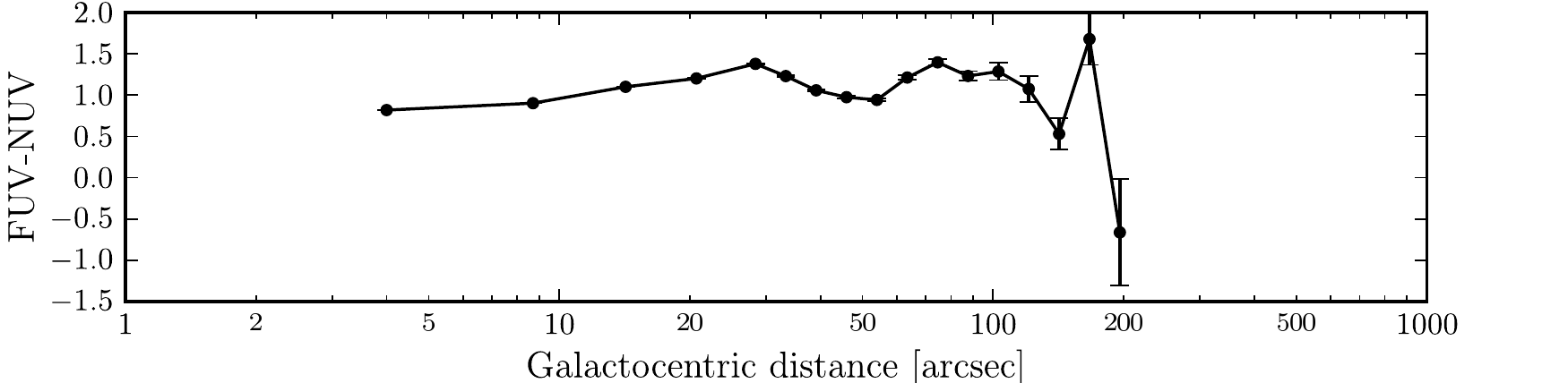}
\caption{\textit{Top panel}: Radial profiles of the \hi \ (blue) FUV (red) and NUV (yellow) emission. The left y-axis shows the UV magnitude and the right y-axis shows the \hi\ surface density. The green-vertical line shows the effective radius of the galaxy. The gray-shaded region indicates the inner \hi \ ring. \textit{Bottom}: Radial FUV-NUV colour profile. The error bars represent the uncertainty in determining the background for both images.}
\label{fig:rp_log}
\end{figure*}

Figure \ref{fig:rp_log} shows the resulting \hi \ and UV radial profiles together. The first peak of the \hi \ profile indicates the location of the inner \hi \ ring $\sim$ 30-80 arcsec from the centre. We find that the slope of the declining FUV profile becomes flatter in this region consistent with the presence of diffuse FUV emission associated with the star formation (see Fig. \ref{fig:images2}). Additionally, the FUV-NUV colour becomes significantly bluer at the same location. This might be because the young stellar populations are dominant in this region. The second peak in the \hi \ profile of Fig. \ref{fig:rp_log} ($\sim$ 120-150 arcsec) is caused by the high column density gas in the western \hi \ arm (see Fig. \ref{fig:images2}). Unlike for the first peak of the \hi \ profile, here the FUV profile does not seem to depart from the general declining trend. However, the FUV-NUV colour seems to become bluer at this location.
The most significant diffuse FUV emission in NGC~4203 comes from a region close to the UV-bright galaxy bulge. The emission from the bulge may contaminate a large region around it because the GALEX PSF is characterized by broad wings out to at least 60 arcsec. To quantify this effect we create a bulge model with a S\'{e}rsic surface brightness profile and convolve it with the FUV PSF. We find that the average contamination from the bulge is $\sim 29.2$ ABmag~arcsec$^{-2}$ between 50 and 60 arcsec. This contribution to the total emission is very low and we ignore this effect in our SFR calculation.
\begin{figure*}
\includegraphics{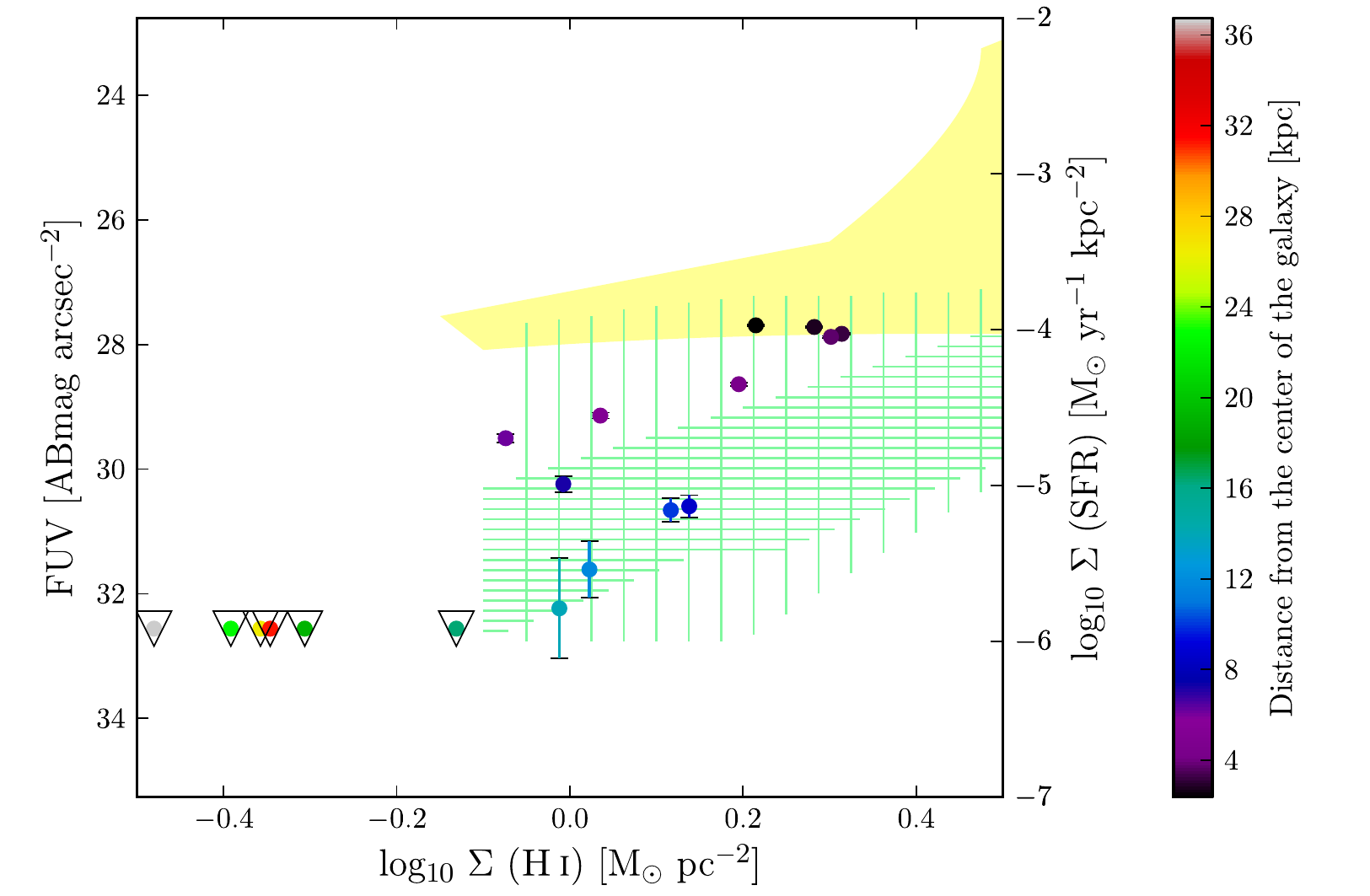}%
\caption{Comparison between \hi \ and FUV surface brightness obtained plotting the radial profiles of Fig. \ref{fig:rp_log} against one another. The triangles indicate \hi \ detected regions with no detectable FUV emission, we estimate the upper limit as equal to the r.m.s. of  the FUV background image. The colour bar shows the distance from the centre of the galaxy. The yellow area shows the relation between \hi \ and SFR in the sample of late-type galaxies of  \citet{2008AJ....136.2846B}. The vertical and horizontal green-shaded areas show the location of dwarf galaxies and outer regions of spirals, respectively \citep{2010AJ....140.1194B}. In this Figure, the left axis shows the FUV surface brightness and the right axis shows the SFR calculated with Eq. \ref{eq:eq3}.}
\label{fig:intermsofradius}
\end{figure*}

In order to take into account the significant 8~$\mu$m non-stellar emission in the inner \hi \ ring (see Fig. \ref{fig:images2}), we derive the SFR radial profiles based on this 8~$\mu$m emission as well as the 24~$\mu$m emission (see Sec. \ref{sec:8mu} and Eqs. \ref{eq:eq4} and \ref{eq:eq5}). We show the resulting SFR radial profiles together with the FUV one in Fig. \ref{fig:spitandfuv}. All profiles show a similar trend as their slope changes at the location of the \hi \ ring (in fact, at 8 $\mu$m the profile increases significantly between 30 and 60 arcsec). The infrared based SFR profiles are up to a factor $\sim$~3 higher than the profile based on the FUV emission alone. This difference is larger than the errors in the photometric measurements and is likely due to infrared emission from obscured star formation. The photometric uncertainty for the \textit{Spitzer} 8 and 24~$\mu$m images is $\sim$ a factor of 2, and this uncertainty is also similar for the FUV measurements \citep{2007ApJ...666..870C, 2007ApJS..173..267S}. However, at present, the depth of the Spitzer images and the small FOV imply that the UV images are the only reliable tracer of SF at larger radius. Table \ref{table:parameters2} shows the SFR efficiencies at the location of the inner \hi \ ring for all 3 tracers as well as their approximate uncertainties. 

\subsection{SFR in terms of \hi \ column density}
\label{sec:sfr_hi}
In Fig. \ref{fig:intermsofradius} we compare \hi \ and SFR surface density by plotting the \hi \ and FUV radial profiles of Fig. \ref{fig:spitandfuv} against one another. In this figure points are colour-coded according to the distance from the galaxy centre. The reason that the track described by our measurements is not monotonic is the result of the two \hi \ peaks (one corresponding to the inner \hi \ ring and the other to the outer \hi \ arm; see Figs. \ref{fig:images2} and \ref{fig:rp_log}) while the UV radial profile decreases monotonically with radius. For comparison, the yellow area shows the spatially resolved relation between \hi \ and SFR surface density inside $r_{25}$ for a sample of nearby (D~$<$~12~Mpc) spirals covering a range of morphologies \citep{2008AJ....136.2846B}. In this inner region, interstellar medium of these galaxies is dominated by molecular hydrogen. The green vertically and horizontally shaded areas show the same relation for the outer parts (i.e., 1-2 $\times \ r_{25}$) of dwarf and spiral galaxies, respectively \citep{2010AJ....140.1194B}. Details about their sample, e.g., the THINGS sample, are given in \citet{2008AJ....136.2563W}. This figure shows that the SF efficiency in the bright \hi \ ring around NGC~4203 is consistent with that of late-type galaxies. Moreover, the SF efficiency in the outer regions of NGC~4203 is also consistent with that of outer regions of spiral and dwarf galaxies.
\begin{figure*}
\includegraphics{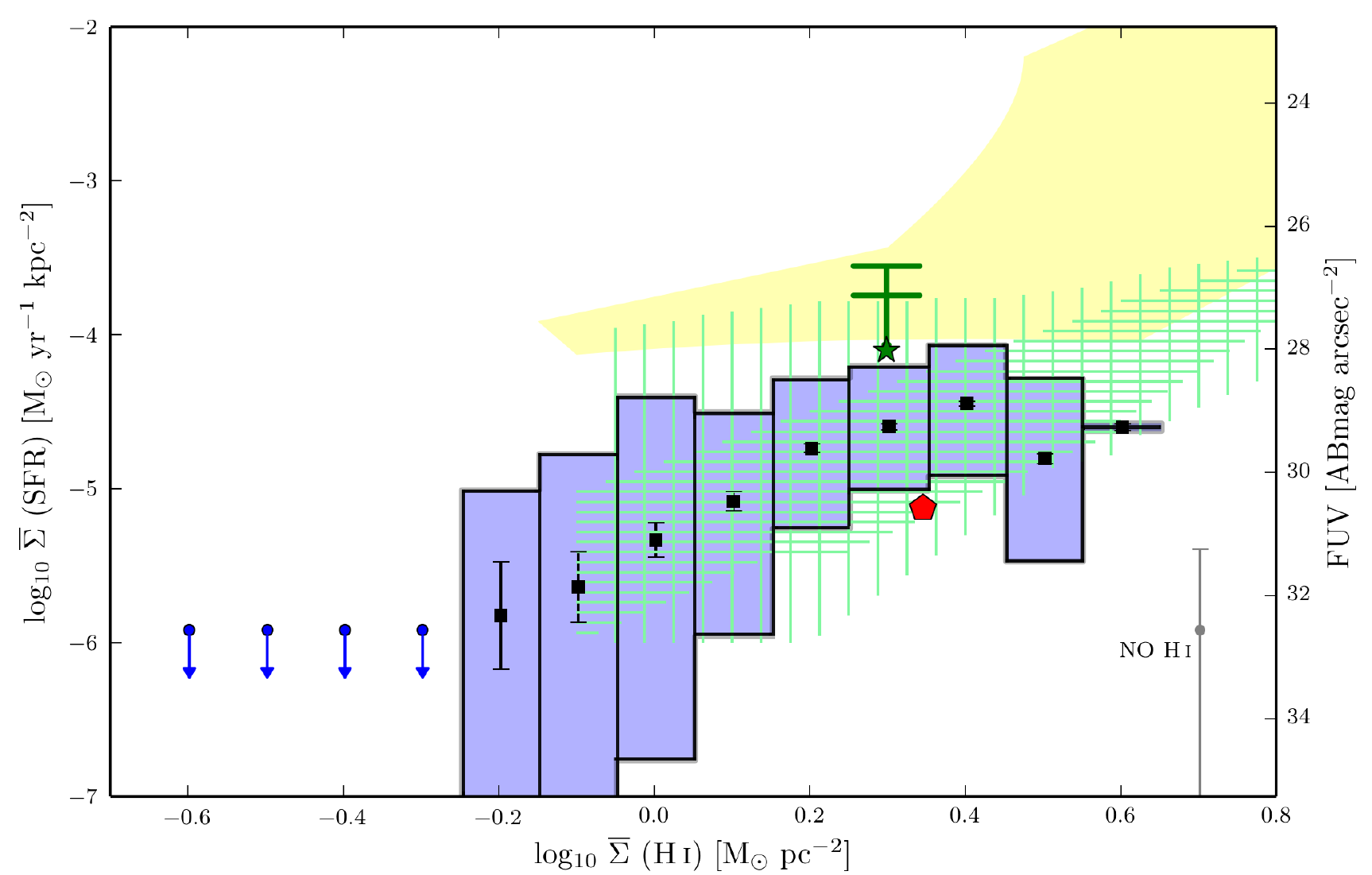}%
\caption{SFR surface density as a function of \hi \ column density based on the pixel by pixel method. Here, the FUV pixels are binned as a function of \hi \ column density. The light blue-shaded area shows the scatter of FUV pixels in the bins. The error bars are similar to those in Fig. \ref{fig:rp_log} and represent the uncertainty in determining the background. The gray error bar on the bottom-right shows the result of same analysis for all the pixels without \hi \ emission. The green asterisk and red pentagon show the inner \hi \ ring and the western \hi \ arm, respectively. The two green lines above the green asterisk represent the SFRs in the inner \hi \ ring that are calculated from Eq.~\ref{eq:eq5}~(for FUV+24~$\mu$m) and \ref{eq:eq4}~(for 8~$\mu$m), respectively. The blue points represent the upper limits, calculated by using the r.m.s. of the FUV background image. The yellow area and green lines are the same as in Figure \ref{fig:intermsofradius}.}
\label{fig:intermsofhi}
\end{figure*}

So far we have compared \hi \ and SFR surface density at fixed radius. Below, we perform a pixel by pixel comparison as implemented by previous authors \citep[e.g.,][]{2008AJ....136.2846B,2010AJ....140.1194B,2012A&A...545A.142B}. To do so, we smooth and re-grid the FUV and infrared images to the resolution and coordinate grid of the \hi \ image. We find this comparison to be characterised by a very large scatter. For this reason, we bin the points along the HI column density axis, and show the average SFR surface density of each bin as well as the scatter in Fig. \ref{fig:intermsofhi}.

This plot shows a weak trend between \hi \ and SFR surface density as well as a large scatter, represented by the blue bars. This large variation of SF efficiency at fixed \hi \ column density is also clear when one compares the \hi \ ring and the western \hi \ arm mentioned above. The green asterisks and red pentagon indicate the average values of SFR surface denstiy and \hi \ column density in these two regions, respectively, and are calculated on the basis of the FUV image. Combining the 24 $\mu$m image with the FUV image, the SFR estimate increases further, as illustrated by the first green line. If we calculate the SFR with the 8~$\mu$m non-steallar emission, we obtain the value that is indicated by the second green bar. As argued above, this strengthens our claim that SF efficiency in the inner \hi \ ring is as high as that of spiral galaxies, while it is much lower in the western \hi \ arm despite the similar gas density. However, we note that even if the SFR is low in the western \hi \ arm, this SFR is still consistent with the outer regions of spirals and dwarf galaxies.

\section{Discussion}
\subsection{Star formation efficiency}
\label{sec:discussion_1}
The previous section demonstrates that at any given \hi \ column density there is a large range of SFR surface densities and, therefore, SF efficiencies in the \hi \ disc of NGC~4203. This is highlighted by the comparison between the \hi \ ring and the western \hi \ arm presented in Fig. \ref{fig:intermsofhi}. There are a number of possible explanations for the difference between these two regions: (i) there may be no direct relation between \hi \ and SF; (ii) the \hi \ column density of the arm might appear to be higher because of projection effects (e.g., the \hi \ arm stretches away from the plane of the disc), and the actual density of the gas could be too low to form stars; (iii) the temperature of the \hi\ in the arm may be higher than in the ring, making it more difficult to form stars; (iv) The metallicity and dust content of the inner and outer regions might be different due to a difference in formation processes; (v) the western \hi \ arm may be stable against gravitational instabilities, preventing the formation of stars.

Both the second and third points would result in a broadening of the \hi \ spectral profile at the location of the \hi \ arm. This is, however, not observed, leaving us with no clear indication in favour of these options. On the other hand, the fourth point seems an attractive explanation. \citet{1999ApJ...510L..95M} observe a quasar $\sim$ 2 arcmin South-East of the galaxy. Interestingly, they detect one of the closest known damped Ly$\alpha$ (DLA) system. This DLA system belongs to NGC~4203, and the metallicity of the gas to be [Fe/H]~=~-2.29~$\pm$~0.1. They also show a number of heavy-element abundances such as [Mn/H]$<$ -0.68 and [Mg/H] $>$-2.4, respectively. The velocity of the lines are identical to that of the \hi \ emission from the disc of NGC~4203 at the position of the quasar within ($\pm$10 kms$^{-1}$). Although the metal abundance values are very low, the depletion of the heavy elements onto small dust grains can affect the observed measurements. \citet{2002A&A...391..407V}, for instance, show that the corrected [Fe/H] abundance of a DLA system can be on average 0.5 dex (in some extreme cases 1 dex) higher than that of observed. Even if we consider 1 dex depletion effect, the metallicity of the outer \hi \ disc remains very low (i.e., [Fe/H] $\sim$ -1.29).

This is likely quite different from the inner \hi \ ring, which is known to contain a considerable amount of dust (Fig. \ref{fig:images2}). In particular, we know that the PAH molecules traced by the 8~$\mu$m emission are vunerable to highly energetic photons in dust poor environments due to low dust shielding \citep[e.g.,][]{2004A&A...428..409B, 2005ApJ...633..871C}, and the PAH emission generally vanishes in regions with metallicity below solar \citep{2005ApJ...628L..29E}. Thus the metallicity in the inner region should be significantly higher than in the outer regions. Consequently, different formation processes for the inner and outer \hi \ regions can explain these different metallicities.

The last possibility is the differences of the gravitational instabilities of the \hi \ gas in the inner and outer regions. For example, star formation can occur in regions where the gas disc is unstable against large-scale collapse. If the gas is stable against it, Coriolis forces counteract the self-gravity of the gas and suppress star formation \citep[][and references therein]{2008AJ....136.2782L}. This might explain the different SF efficiencies. However, in order to understand this effect, we need further investigation.

\subsection{Origin of the \hi \ gas}
\label{sec:discussion_2}
There are different scenarios that could explain the origin of the \hi \ disc of NGC~4203. One possibility is a high-angular-momentum, gas-rich major merger. Such a merger could result in the formation of a large gas disc \citep[e.g.,][]{2002MNRAS.333..481B}. At the same time, it would not necessarily trigger the central burst of SF usually thought to occur during mergers  \citep[e.g.,][]{2007A&A...468...61D,2008A&A...483...57S}, as evidenced by the old stellar populations in the centre of NGC~4203. If this scenario is correct, we should able to see a residual from the major merger close to the centre, or a distortion in the outer stellar halo of NGC~4203 in the deep optical image. However, we do not detect any of these features in Fig. \ref{fig:images3}. Another problem with the major merger scenario is the extremely low metallicity of the outer disc which may rule out that the \hi \ once belonged to a large galaxy. 

The other possibility is that the gas has been accreted from the inter-galactic medium. The disturbed appearance of the \hi \ disc may be explained by some minor interaction with a satellite. Indeed, as explained in Sec. \ref{sec:propertiesHI}, there is a nearby dwarf satellite, which is most likely interacting with our galaxy. Although this dwarf is positioned at the end of one of the \hi \ arms of NGC 4203 (Fig. \ref{fig:images}), we do not detect any \hi \ emission at that position and, therefore, rule out that it is the donor of the \hi \ gas.

However, there is another possibility that this gas can be captured from another gas-rich dwarf galaxy in a single encounter. Since the \hi \ disc is almost regular, this event would happen at least a couple Gyr ago, making it difficult to detect. This is consistent with the conclusion of \citet{1988A&A...191..201V} based on a study of the environment around the galaxy.

In conclusion, accretion from the inter-galactic medium would explain the low metallicity of the outer disc, while the inner \hi \ ring may have been enriched of metals by stellar mass loss in the galaxy. Moreover, the disturbed appearance of the outer disc can be explained by an interaction with the satellite visible in Fig. \ref{fig:images} and \ref{fig:images3}.

We have long known that a large fraction of all ETGs host low levels of on-going or recent SF. This SF activity is detected in their centre using optical spectroscopy \citep[e.g.,][]{2000AJ....119.1645T} or UV photometry \citep[e.g.,][]{2005ApJ...619L.111Y}. Similar to extended-UV discs in late-type galaxies, star formation can also be detected at large radius in early-type galaxies using UV and deep optical imaging \citep[e.g.,][]{2010ApJ...714L.290S, 2012ApJ...755..105S, 2012ApJ...761...23F, 2015MNRAS.446..120D}. In a few cases it has been possible to establish a link between the presence of \hi \ and SF at the outskirts of ETGs \citep[e.g.,][]{2009MNRAS.400.1225C, 2009AJ....137.5037D, 2010ApJ...714L.171T, 2014MNRAS.440.1458D}. Here we explore such link for NGC~4203, and in this section we quantify the impact that the SF hosted within the \hi \ disc has on the morphology of the host.

Table \ref{table:parameters2} shows the result of our SFR calculation based on the various tracers discussed above, and performed separately for the inner \hi \ ring and the remaining, outer \hi \ disc. In particular, the inner \hi \ ring hosts a significant amount of dust-obscured SF, and the SFR is best estimated using a combination of 24~$\mu$m and FUV emission.

The inner \hi \ ring and outer \hi \ disc are very different from each other. The former contains only $\sim$10 percent of the total \hi \ mass but hosts $\sim$97 percent of the total SF. The latter hosts the remaining $\sim$90 percent of the \hi \ mass but only $\sim$3 percent of the total SF. As a result the \hi \ depletion time is much longer in the outer disc ($\sim10^3$ Gyr) than in the inner \hi \ ring ($\sim$5 Gyr). If nothing happens to the galaxy, the \hi \ disc could survive for hundreds of a Hubble time at the current SFR.

Overall, the detected amount of SF has a negligible effect on the stellar disc growth and, therefore, the morphology of NGC~4203. Based on the total stellar mass reported by \citet{2013MNRAS.432.1862C} and on the B/D ratio given in \citet{2013MNRAS.432.1768K}, the mass of the stellar disc is $\sim$2.7 $\times 10^{10}$ M$_{\odot}$. At the current SFR, this mass would grow by a mere 0.1 percent in 1 Gyr. The star formation rate per unit stellar mass in the stellar disc is of just $\sim$ $10^{-12}$ yr$^{-1}$.

If we suddenly convert all the gas in the outer \hi \ disc of NGC~4203 to stars, the R band surface brightness of this region would be 26.8 mag arcsec$^{-2}$. This is $\sim$ 2 magnitude lower than Malin 1, a giant low surface brightness galaxy that has a low surface brightness and very large \hi \ disc \citep{2010A&A...516A..11L}. In conclusion it is unlikely that NGC~4203 will become a galaxy similar to Malin 1.

\section{Conclusions}
\label{sec:conclusions}
NGC~4203 has an \hi \ disc that consists of two separate components. The first component is the inner star forming \hi \ ring. This ring contains metal-rich gas and a large amount of dust. The SF efficiency in the \hi \ ring is comparable to that of inner regions of spiral galaxies and the \hi \ depletion time is of the order of a few Gyr. The second component is the outer \hi \ disc, where the metallicity is likely to be low. The SF efficiency in this disc is much lower than that of centre of spirals even when the \hi \ column density is high (e.g., in the western \hi \ arm) and the \hi \ depletion time is of 100's of the Hubble time. However, the SF effieciency in the outer \hi \ disc is still consistent with that of the outer regions of spiral galaxies and dwarfs.

These two components of the \hi \ structure may be formed in different ways. On the one hand, the \hi \ ring is most likely enriched by stellar processes within the galaxy. On the other hand, the outer \hi \ disc might be accreted from the low metallicity IGM because the outer disc is formed with metal-poor, high column density gas and we cannot find a suitable donor. Subsequent interaction with satellites, which is possibly on-going, may explain the current distorted morphology of the \hi \ disc. 

We also detect spiral-like structure in the 8~$\mu$m and deep $g'-r'$ images. These spiral-like structure is extending to 3 effective radii ($\sim$ 6kpc), and, because of its clumpiness, is likely organized by the star formation process.

Finally, despite the large \hi \ reservoir and the detection of some SF, the stellar disc of NGC~4203 is growing at a very low rate and the morphology of the galaxy is unlikely to change significantly in the foreseeable future.
\section*{Acknowledgments}
MKY acknowledges a Ph.D. scholarship from The Council of Higher Education of Turkey and is supported by the University of Erciyes.

We would like to thank the GALEX team for their great work to make the data public and available. This work uses observations made with the NASA Galaxy Evolution Explorer. GALEX is operated for NASA by Caltech under NASA contract NAS5-98034. Part of this work is based on observations made with the Spitzer Space Telescope, which is operated by the Jet Propulsion Laboratory(JPL), California Institute of Technology under a contract with NASA. This research has made use of the NASA/IPAC Extragalactic Database (NED) which is operated by JPL, Caltech, under contract with NASA.

\end{document}